\documentclass[twocolumn,preprintnumbers,amsmath,amssymb]{revtex4-2}
\usepackage{graphicx}
\usepackage{color}
\usepackage{caption}
\usepackage{subfigure}
\begin{document}

\title{A structural determinant of the behavior of water at hydration and nanoconfinement conditions}

\author{Nicol\'as A. Loubet$^1$}
\author{Alejandro R. Verde$^1$}
\author{Gustavo A. Appignanesi$^{1 *}$}

\affiliation{
$^1$ INQUISUR, Departamento de Qu\'{i}mica, Universidad Nacional del Sur (UNS)-CONICET, Avenida Alem 1253, 8000 Bah\'{i}a Blanca, Argentina\\
* Corresponding author: appignan@criba.edu.ar\\
}

\date{\today}

\begin{abstract}

The molecular nature of the phases that conform the two-liquids scenario is elucidated in this work in the light of a molecular principle governing water structuring, which unveils the relevance of the contraction and reorientation of the second molecular shell to allow for the existence of coordination defects in water's hydrogen-bond network. In turn, such principle is shown to also determine the behavior of hydration and nanoconfined water while enabling to define conditions for wettability (quantifying hydrophobicity and predicting drying transitions), thus opening the possibility to unravel the active role of water in central fields of research.

\end{abstract}


\maketitle

 \section {Introduction} Liquid water is not only a fascinating and challenging system in itself\cite{gallo_chemrev,angell_02,angell_04,Debenedetti_book,Tanaka00,review_epje_1}, but its behavior is also determinant for a huge number of phenomena that are central to our life and that impact in fields that range from biology to materials science\cite{chaplin, hassanali_chem_rev,ball,review_epje_2,debenedetti,berne,AFlibro,water_m1,water_m4,water_m8,marcia,bordin,paolaV4,graphene,cavities,membrane,martelli_franzese,
hydrophobicity1,hydrophobicity2,hydrophobicity3,bizzarri,short_time,identifying,high-low,proteins1,proteins2}. In turn, water is known to present several structural, thermodynamic and dynamic anomalies which seem to stem from its ability to coexist in two competing different liquid forms (low density liquid, LDL, and high density liquid, HDL), as described by the two-liquids scenario\cite{ball,chaplin,gallo_chemrev,review_epje_1,review_epje_2}. This description emerged from the proposition of the existence of a second liquid-liquid critical point, which has recently received strong validation\cite{gallo_chemrev,review_epje_1,Poole92,LLCP_T1,LLCP_T2,Biddle,LLCP_T3,Pablo-Francesco,LLCP_E1,LLCP_E2,LLCP_E3,Kim}. Some of these anomalies are pivotal for making water irreplaceable by normal liquids in contexts where its plays an active role\cite{chaplin,hassanali_chem_rev,ball,review_epje_2}. Remarkably, water is considered to be an indispensable condition for the existence of life\cite{ball,chaplin}; indeed, when we search for the possibility of life outside planet earth, what we do in fact is to look for water\cite{ball}. Hydrogen bonds (HB), the relatively strong and tetrahedrally-directed intermolecular interactions that promote water structuring, are essential for the anomalies that distinguish it from simple liquids\cite{ball,chaplin,gallo_chemrev,review_epje_1,review_epje_2}. This is precisely the line of reasoning that will be followed in the present work to reveal a molecular principle (embodied in a structure-energy mutual interplay) underlying the structuring of water's HB network. More significantly, such principle will enable us to elucidate the molecular nature of the phases that conform the prominent two-liquids scenario. Specifically, we shall show that while the low density phase is a result of water's unusual ability to expand its second molecular shell in order to improve HB coordination and build an extensive HB-network, contraction and reorientation of the second molecular shell is demanded to partially compensate the HB coordination defects that characterize the high density phase.

At hydration conditions, that is, when water is in contact with a solute, a biomolecule or any other material, a fraction of the water-water interactions should be replaced by the surface in order to produce wetting\cite{chaplin,water_m1,epi,AFlibro}. In some cases where this compensation is provided by means of water-surface HBs or other significant water-surface interactions, a good wetting is achieved, and the surface is termed as hydrophilic\cite{chaplin,debenedetti2,review_Garde}. However, the high reluctance of water to lose HB-coordination\cite{water_m1,epi,AFlibro} poses strong constraints to surfaces that interact weakly with water so that this compensation cannot be fulfilled, thus resulting in a poor wetting or hydrophobic behavior \cite{chaplin,debenedetti2,review_Garde}. Nevertheless, a full comprehension of the relative hydrophobicity of the different surfaces in molecular terms is still lacking. An example in this regard is the unexpected high hydrophilicity of graphene-like surfaces\cite{graphene-exp,graphene-hydrophilic,Fluid-Phase-Equilibria}, which we shall hereby consider. Additionally, the local geometry (as in the case of cavities, pores, etc.) might conspire with the possibility for water to retain a proper HB environment, even to the point to promote drying in certain conditions\cite{Fluid-Phase-Equilibria}. Indeed, nanoconfinement affects dramatically some properties of water and, if an appropriate HB network cannot be established, the liquid becomes thermodynamically unstable\cite{debenedetti} as, for example, in the case of the hydrophobic collapse that promotes protein folding or in the case of biomolecular binding, where the removal of labile water by a ligand has been established as a principal driving force\cite{water_m1,epi,AFlibro,HBPLOS,membranes}. However, water indeed penetrates certain subnanometric cavities, as it happens in small-radius nanotubes, a fact long ago regarded as ``surprising'' on energetic grounds\cite{HummerCNT} and which we shall hereby revisit. The above-expounded scenario reveals the fact that the understanding of the principles of water hydration, including the development of a theory of wettability and hydrophobicity at the nanoscale, is imperative both from a fundamental and an applied standpoint. Notwithstanding, this picture is still far from being complete \cite{Giovambattista,debenedetti}. In this context, we shall demonstrate that the same molecular principle that lies at the heart of the two-liquids scenario for bulk water is also operative in situations of hydration and nanoconfinement, thus extending the validity of the two-liquids description to these realms and unraveling its role in ruling hydrophobicity and wetting.

The present work will provide additional information regarding the nature of the interactions that define the  previously introduced \cite{v4s} index while extending its application as a hydrophobicity measure. Thus, we shall study the  way in which the second shell compensates energetically the missing HB as it contracts towards the first shell, by quantifying the role of the different neighbors. We shall also study the effect of temperature on wetting by considering hydrophobic and hydrophilic behavior at a temperature range that covers from ambient temperature to temperatures within the supercooling regime. Additionally, we shall explore the sensitivity of the hydrophilic behavior of graphene on the strength of the water-wall interactions. Finally, we shall perform a full study of the index distributions for carbon nanotubes with variable C-O interactions in order to reach a deeper understanding of which interactions are responsible for its wetting response and of the extent of such effect.

\section {Methods} In this work we use a new simple structural index to characterize water structure in terms of directional (tetrahedral) intermolecular interactions\cite{v4s}. This parameter is called $V_{4S}$, since it computes the different energetic contributions at four sites tetrahedrally arranged around each water molecule. $V_{4S}$ is calculated for any given central water molecule as follows\cite{v4s}: First, the H-O-H angle is opened until it reaches the value of a tetrahedral angle, $arccos(-1/3)$ (this step is not necessary for models in which the H-O-H angle is tetrahedral, like SPC/E). Subsequently, we define two tetrahedral sites located at 1\AA\ away from the oxygen in both directions and geometrically complete the two other tetrahedrally oriented sites (also located 1\AA\ away from the oxygen). In this way, the four points/sites of a perfect tetrahedron are determined. Thus, this method partitions space in four regions as dictated by the tetrahedral arrangement of the four sites: We select all the neighboring water molecules that are within a distance of 5\AA\ from each of the four sites. Then, we assign each of these molecules to their closest site. We chose 5\AA\ in order to include roughly both the first and second neighbor shells of the central molecule in the case of bulk water. In turn, for the situations in which we consider the interaction with a surface, we also include all the heavy atoms of the surface that lie within the spheres of radius 5\AA\ centered at the four tetrahedral sites of the water molecule of interest, and assign each of them to their closest site. Then, for each tetrahedral site we add up the contributions of all the pair-wise interaction potentials between the central molecule and all its assigned water molecules, and also all its assigned heavy atoms in the case of surfaces, taking into account  Lennard-Jones and the Coulomb interactions. With this procedure, we are left with four potential values, one for each point/site of the tetrahedron. These are ordered so that $V_{1S}$ is the most interacting (most negative value), and $V_{4S}$ is the weakest interacting. As for other previous water structural indicators like $V_4$\cite{v4,v4T2}, both for bulk water and for the water molecules in contact with the different systems studied, we calculated the index for minimized configurations (inherent structures, IS) by using the steepest-descent method to avoid the blurring effect of thermal energy\cite{v4,v4T2}. We note that the whole box is minimized (keeping restraints when corresponding) so that each instantaneous configuration from the molecular dynamics is mapped into its corresponding IS. In fact, the effect of the energy minimization is to properly discriminate HB coordination defects (involving effective HB rupture) from merely deformed structures, since certain water molecules which would be classified as defects (unstructured) indeed belong to structured basins of attraction in the potential energy surface (the process restores the deformed HB). 
We note that the $V_{4S}$ index is suitable for generic contexts, since the energy contributions we consider at each of these four sites could arise either from the interaction of the oxygen of the central molecule with its set of neighboring water molecules and/or with the heavy atoms of any other system. Thus, it could be applied both in bulk conditions or at interfaces or nanoconfinement. 

In order to try to understand the structural basis of hydration, we decided to apply our new structural indicator to model systems: self-assembled monolayers (SAMs) functionalized to act as hydrophobic or hydrophilic surfaces\cite{review_Garde,graphene-hydrophilic,Fluid-Phase-Equilibria}, graphene sheets and carbon nanotubes, CNT. Again, we note that previous structural indicators for water would not be appropriate in this context\cite{COMMENT-PRL}.

\begin{figure}[!]
\resizebox{0.5\textwidth}{!}{%
\includegraphics{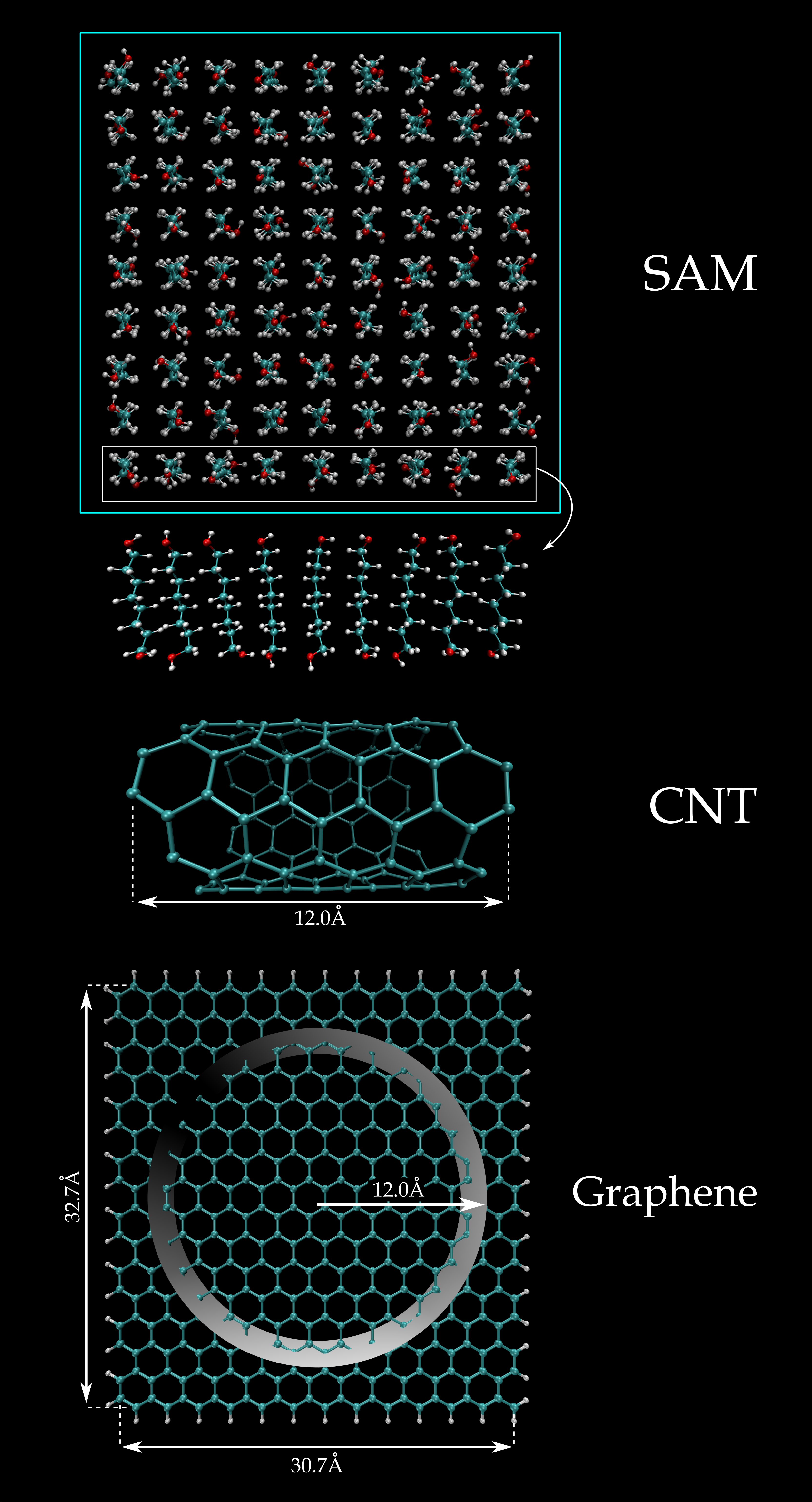}
}
\captionsetup{font=small} 
\caption{Schemes of the systems used ($SAM-OH$, CNT and grapehene). For the sake of clarity we only show the systems and not the water molecules included in each case.}
\label{fig1}
\end{figure}

The graphene sheet consisted in a perfect honeycomb graphite-like sheet of 416 carbon atoms with terminations in hydrogen atoms (we employed a graphene sheet of 32.7 x 30.7\AA\ and analyzed a cylinder of 12\AA\ at the middle of the system) while the carbon nanotube (CNT) studied consisted in  a 112-carbon atoms CNT of 7.3\AA\ diameter (we employed a CNT with a length of 12.0\AA\ immersed in water including only carbon, that is, not terminated in hydrogens). In both cases the systems were solvated within large boxes of TIP4P/2005 or TIP3P water molecules. In the case of the CNT, the TIP3P water model was chosen to compare with the results of \cite{HummerCNT}. Using the GROMACS package\cite{gromacs} and the General AMBER force field (GAFF), these systems were stabilized in NVT conditions at 300K for several structural relaxation times ($\tau_\alpha$), and subsequently stabilized in NPT conditions at 1bar and the same temperature. Finally, 40ns MD simulation were carried out, printing configurations every 0.4ps. In all the stabilizations and dynamics, the initial position of the system was restrained (center of the box). In the case of the normal carbon nanotube, the interaction parameters used were $\epsilon_{C-O}=0.478369kJ~{mol}^{-1}$ and $\sigma_{C-O}=3.27521$\AA. In addition to this dynamics, 16 other 30ns dynamics were performed each with configurations every 6ps, linearly varying the interaction parameters from $\epsilon_{C-O}=0.478369kJ~{mol}^{-1}$ and $\sigma_{C-O}=3.27510$\AA to $\epsilon_{C-O}=0.228720kJ~{mol}^{-1}$ and $\sigma_{C-O}=3.44154$\AA.  In the case of the SAMs\cite{graphene-hydrophilic,Fluid-Phase-Equilibria}, we aligned 81 chains ($9 \times 9$, chain separations of $4.35$ \AA ) of n-decane ($CH_3-(CH_2)_{8}-CH_3$) in a parallel way so that their heads adopted a square-like arrangement facing water (we worked with these square-like arrangements since we only wanted to use generic hydrophobic and hydrophilic systems and we employed positional restraints in order to fix the position of the system in space). The chain heads were functionalized in two different ways: they were ended with methyl groups (for the $SAM-CH_3$) or hydroxyls ($SAM-OH$). The systems were solvated with 10654 TIP3P water molecules in an orthogonal cubic box with periodic boundary conditions (that extended more than $20$ \AA\ away from all the monolayer faces).
For all dynamics simulations, electrostatic pair interactions were used for all atoms, while Lennard-Jones interactions were applied to heavy atoms, with a cut-off distance of 9\AA. The NVT and NPT stabilization and production phases were thermalized using the Berendsen thermostat\cite{berendsen}, with a coupling constant of 0.1 ps, except for the case of the SAMs that were simulated with the AMBER10 molecular simulation suite\cite{amber} with a 2 fs time step (we used the GAFF and FF99SB force fields) and all calculations were performed in the NPT ensemble with a Langevin thermostat. For NPT stabilization and production, the pressure was maintained at 1 bar using the Parrinello-Rahman\cite{prbarostat} barostat for all cases, with a coupling constant of 2.0 ps. Fig.~\ref{fig1} depicts some of the systems under study.

\section {Results and discussion} 
\subsection {Bulk water}
To begin, let us pose the following ansatz on the molecular structuring of the HB network of liquid water: Provided its strong reluctance to lose its tetrahedrally-directed coordination, anytime a water molecule resigns a HB, certain directional energetic compensation would be demanded to avoid the otherwise uncompensated partial molecular charges\cite{AFlibro,epi,water_m1}. For bulk water, we shall show that this requisite is met by a fine-tuned interplay between the first and second neighbor shells, in line with the two-liquids scenario. In turn, in the case of surface hydration and water nanoscale confinement, this behavior should also be locally mimicked by the interacting material in order to ensure hydrophilic wettability, as we shall also demonstrate. Thus, this operational principle will be first validated in bulk water by a simple structural metric which will then be applied to model surfaces and nanoscale confinement to exhibit its potential to characterize hydrophobicity and predict filling and drying transitions. 
Specifically, given the relevant role of directional (tetrahedral) intermolecular interactions in water structuring, we employ a new simple index to characterize water structure called $V_{4S}$\cite{v4s}. To define it (see Methods for more details), for each molecular dynamics configuration of the system, we use its inherent structure (that is, we minimize potential energy to get rid of the blurring effect of thermal fluctuations). Centered at the oxygen of a water molecule whose index we want to calculate, we define a perfect tetrahedron whose vertices are located at 1\AA\ from the center, thus defining four sites. Two vertices of the tetrahedron are placed in the directions of the H atoms (in case of water models whose H-O-H angle is not perfectly tetrahedral, we need to open the angle accordingly). Then we define spheres of 5 \AA\ radius centered at each site and we assign each one of the heavy atoms within these spheres (the oxygens of other water molecules or any heavy atoms that might be present in non-bulk conditions) to their closest site of the tetrahedron. At each site, we then algebraically sum the pairwise interactions (Lennard-Jones and Coulombic) of the central oxygen with all the heavy atoms assigned to such tetrahedral point, in order to determine the potential energy value of such site. Then we order the four potentials starting from the most intense, to get $V_{1S}$, ..., $V_{4S}$, so that $V_{4S}$ is the least interacting tetrahedral site. It is relevant to note that this index is valid not only for bulk water but that it has been specifically devised to be suitable in contexts where the application of other structural indicators would produce artifacts, as we have shown previously\cite{COMMENT-PRL}.

\begin{figure}[!]
\resizebox{0.5\textwidth}{!}{%
\includegraphics{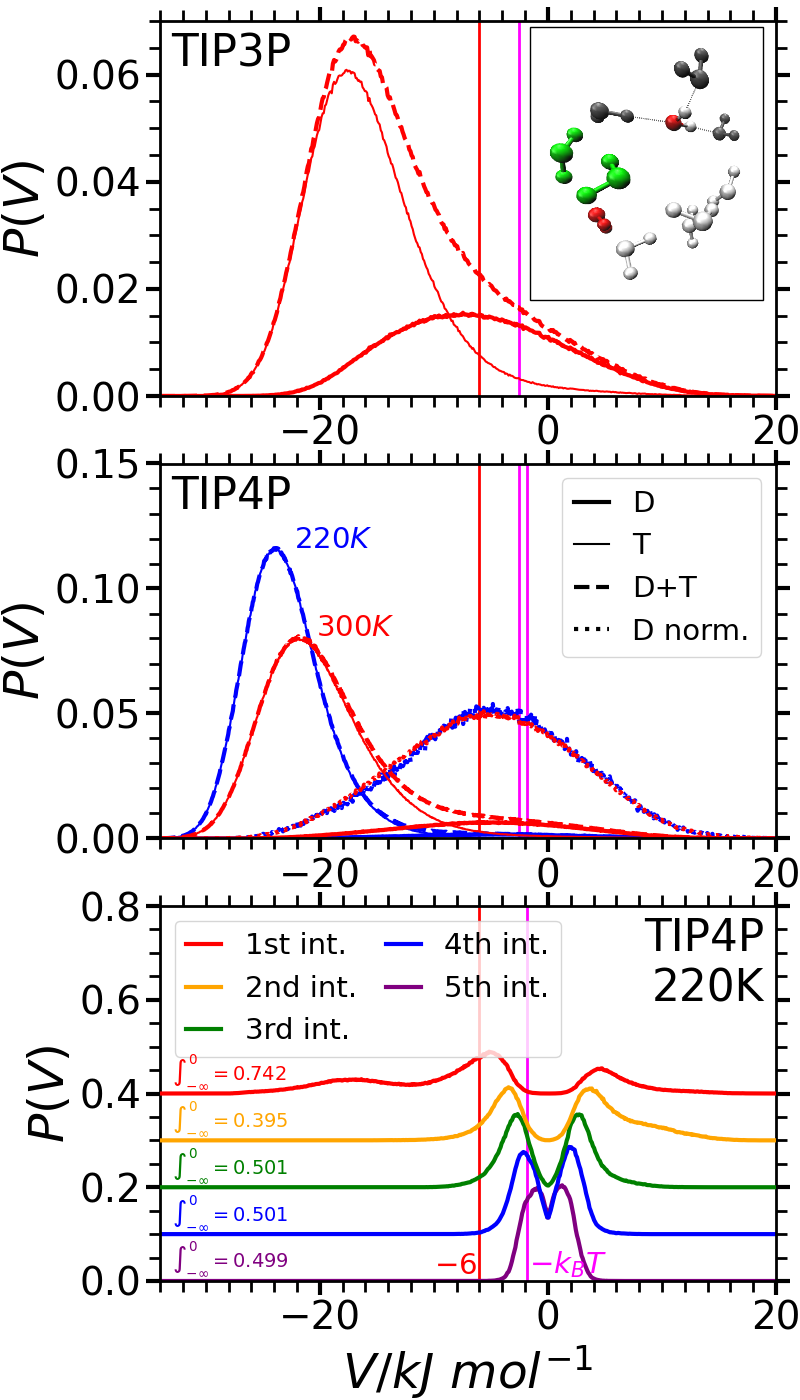}
}
\captionsetup{font=small} 
\caption{(top) Distributions of the $V_{4S}$ index calculated at the inherent dynamics for TIP3P water at 1 bar for T=300K (dashed curve). We also discriminate the contributions of the T molecules (low local density, with good tetrahedral HB-coordination; thin solid line) and D molecules (3-fold HB-coordinated and high local density; thick solid line). The inset shows an example of the environment of a D central molecule, where we indicate in gray the molecules that form HB with it, in green the other attractive ones (more than $k_B T$, where $k_B$ is Boltzmann's constant), in red the repulsive ones and in white the ones whose interaction with the central one is almost null; (medium) $V_{4S}$ distribution for TIP4P/2005 water model at P=1bar and two temperatures: T=300K and T=220K. For both temperatures, we also include normalized (unitary area) curves for the D molecules; (bottom) Distributions of the first five water molecules that contribute to the $V_{4S}$ value for TIP4P/2005 at T=220K, ordered from highest to lowest interaction intensity (we shift the curves for better visualization). We also integrate the attractive part of the energy distributions and indicate each of them over their corresponding curve.}
\label{fig2}
\end{figure}

Fig.~\ref{fig2} displays the distribution of the $V_{4S}$ index for bulk water (dashed lines), which exhibits signs of bimodality. Such behavior is clearly evident (see solid lines) if we additionally classify the water molecules in terms of the previously defined $V_4$ index which, for a given central molecule, finds the four water molecules with which it interacts more strongly and picks the value of the less intense of such interactions\cite{v4,v4T2}. This parameter has been successful in dissecting water molecules into LDL-like (called T molecules) and defects (D molecules)\cite{v4,v4T2}. T molecules are characterized by a good four-fold tetrahedral HB-coordination in their first shell (interactions stronger than 12 kJ/mol, compatible with good quality HBs), while the D ones present three-fold HB coordination and are responsible for the existence of the HDL state\cite{v4,v4T2}. Specifically, in Fig.~\ref{fig2} (top) we discriminate the relative contributions of the T (solid thin line) and D (solid thick line) molecules to the $V_{4S}$ distribution for TIP3P water model at 300K temperature (in Fig.~\ref{fig2} (medium) we provide the corresponding results for TIP4P/2005 water model, also including the case for a temperature of T=220K, well within the supercooled regime, thus presenting a very low fraction of D molecules). From Fig.~\ref{fig2} (top), we find that the T molecules peak to the left (at around  $V_{4S} =-17 kJ/mol$), while the D ones are broadly distributed to the right with a mean value of $V_{4S} = -6.08$ kJ/mol. TIP4P/2005 (Fig.~\ref{fig2} (medium)) exhibits a similar behavior, with the peak for the T molecules located a bit to the left as compared to TIP3P, and even more to the left for the case of T=220K, thus speaking of a better structuring behavior. This is expected since TIP3P presents an abnormally low melting point (also, at T=300K, the fraction of D molecules for TIP3P is much larger than that for TIP4P/2005, which is relatively low). In turn, the peak for the D molecules is located at roughly the same position for both water models and behaves as temperature-insensitive, as can be learned from the results for TIP4P/2005 (in Fig.~\ref{fig2} (medium) we show both the curves for the contribution to $V_{4S}$  of the corresponding fractions of D molecules and also their normalized versions, that is, making them full probability distributions, with unitary area). At this point, it is relevant to note that while the value for the peak of the D molecules is less intense than that for the T molecules, it is still quite attractive. We note that in a previous work\cite{v4} we have shown that the radial distribution functions (RDF) of T and D molecules are very similar to the experimental LDA and HDA, respectively\cite{HDA-LDA}. For example, if we integrate the RDF of the D and T molecules up to 4\AA\\ for TIP4P/2005 at T=210K we find that the former yield a value 13.6\% larger than the latter (that is, the T molecules are indeed less dense or more expanded). The corresponding difference between HDA and LDA is 19.5\% when we perform the same calculation on the experimental data of\cite{HDA-LDA}. We also wish to stress that when we speak of a low and a high density pahse there is the need to consider multi-molecular descriptions\cite{v4,v4T2,v4s}. Indeed, studies of the spatial distribution of T and D molecules show that the T molecules conform a low density phase but that the high density state incorporates D molecules together with their environments. This implied to go beyond the single-molecule description to a multi-molecular one and enabled to find isofractions at a temperature consistent with the maxima in themodynamic response functions as expected by the Widom line implied in the two-liquids scenario\cite{v4,v4T2,v4s}.

From one side, these results show that it is evident that T molecules leverage water's unusual ability to expand its structure to gain energy: an expansion of the second neighbors shell allows the first one to enjoy a full 4-fold tetrahedral HB-coordination with the central molecule. Thus, the $V_{4S}$ value for the T molecules is dominated by the energy of a first-shell HB with virtually no contribution of the second shell. D molecules, in turn, imply a broken HB, since second-shell molecules have penetrated the intershell gap (thus increasing the local density) and, hence, their first shell comprises only 3 HB-coordinated neighbors. However, as already recognized\cite{HDL}, the largely overlooked HDL state is far from being unstructured. From the results of the $V_{4S}$ distribution, it is now evident that the closer second-shell molecules tend to arrange themselves in dipolar orientations to compensate the partial molecular charge of the central molecule at the tetrahedral site of the lacking HB (this can be achieved by many local arrangements; see an example shown in the inset of Fig.~\ref{fig2} (top)). In this sense, the central molecule recovers part of the HB-loss, as evidenced by the peak at an attractive value at around -6 kJ/mol. We note that while one second-shell molecule tends to contribute substantially, sometimes two or three second-shell molecules are also engaged in such an energetic compensation. In Fig.~\ref{fig2} (bottom), we also provide the distributions of the first five water molecules that contribute to the $V_{4S}$ value for TIP4P/2005 at T=220K, ordered from highest to lowest interaction intensity. It is evident that the first interaction is the one that contributes the most, being generally attractive (we include in the plot the value obtained by integrating the attractive part of the distributions, that is, the negative region). The second interaction is a bit repulsive on average and the other ones display an indistinctive tendency. However, in all cases except for the fifth one, a peak showing a slight attraction (more intense than $k_B T$, also marked in the plot) is evident. This shows that while the first interaction tends to be the one that dominates the energetic compensation, more than one molecule is sometimes involved. 

In summary, our results show that, while the expanded LDL state is dominated by first-shell energetics, the high local density molecules display an enhanced contribution of the partially contracted second-shell. This interplay between structure and energetics (stemming from the directional nature of the HB interactions) evidences the unique ability of water to present, unlike simple liquids, a marked structural plasticity by taking advantage of two opposed structural/energetic trends. Hence, a first interesting lesson that emerges from these results is that the existence of the LDL and HDL molecular states in bulk water might result from this fine-tuning in the arrangement of the two first molecular shells in order to satisfy the energetic constraints of the previously enunciated ansatz.

\begin{figure}[h!]
\resizebox{0.5\textwidth}{!}{%
\includegraphics{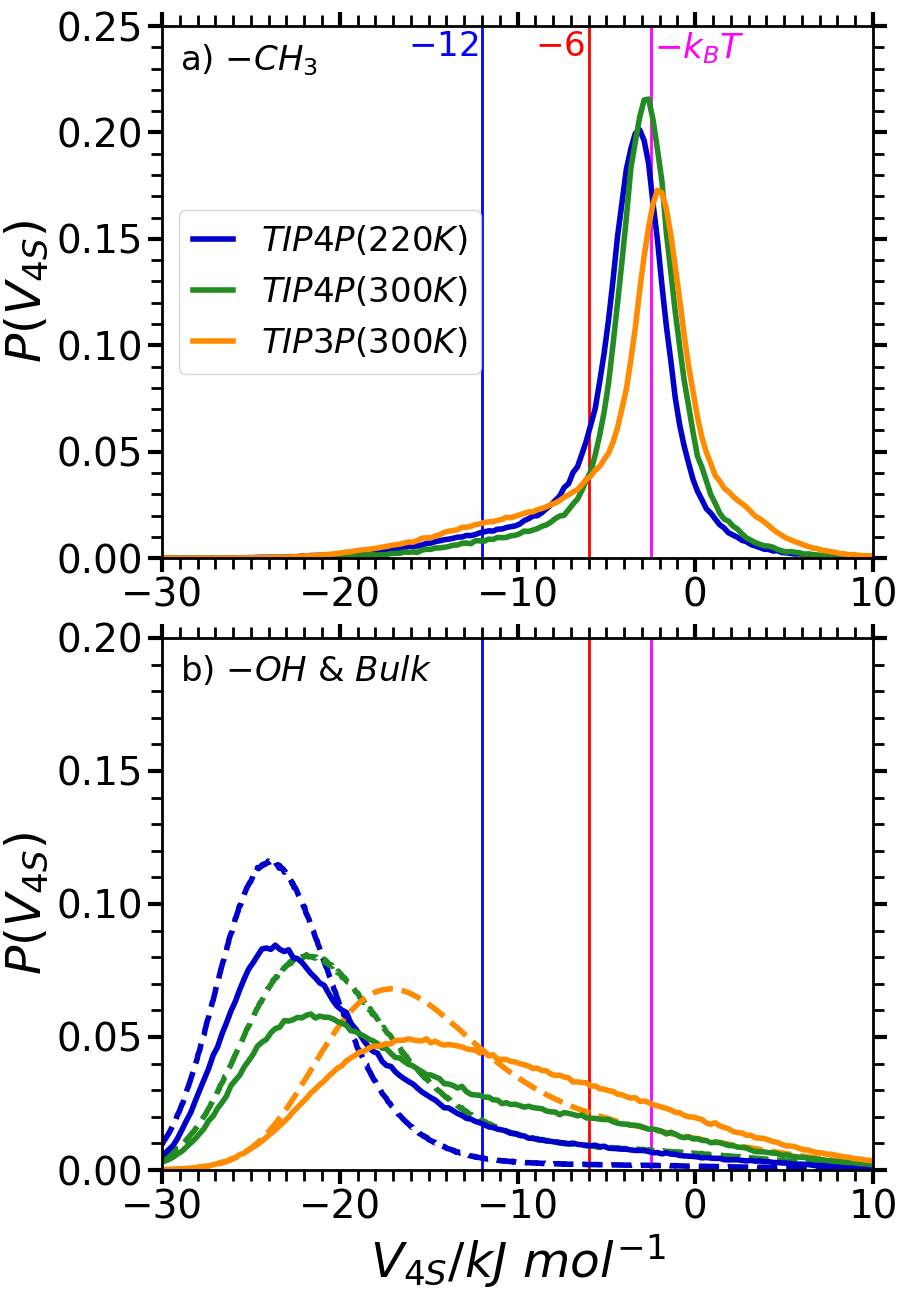}
}
\captionsetup{font=small}
\caption{a) Distributions of the $V_{4S}$ index for the water molecules close to the $SAM-CH_3$. We show the case of TIP3P water at 1 bar for T=300K together with TIP4P/2005 at 1 bar and T=300K and T=220K. b) Idem for the water molecules close to the $SAM-OH$, also including the corresponding $V_{4S}$ distributions for bulk water (dashed lines).}
\label{fig3}
\end{figure}

\subsection {Hydration water}
In order to try to understand the structural basis of hydration, we decided to investigate a model system, a self-assembled monolayer (SAM), functionalized to act either as a hydrophobic or as a hydrophilic surface\cite{review_Garde,graphene-hydrophilic,Fluid-Phase-Equilibria}. In Fig.~\ref{fig3} we show results of the $V_{4S}$ parameter calculated for the hydration water molecules in both cases. We consider water molecules that are in contact with the surface (located closer than 3.5\AA\ from the surface, in order to include the first peak of the water density profile normal to the surfaces). We show results for TIP4P/2005 and TIP3P water models at a temperature of T=300K (we also include the case of T=220K for the former). In the case of the  hydrophilic SAM ($SAM-OH$), the resulting distributions are very similar to that for the corresponding bulk water case (which we also include for comparison). In this case, the water molecules tend to form three water-water HBs and an energetically analogous fourth one is formed with the surface hydroxyls, thus creating an environment equivalent to that of the 4-fold coordinated bulk molecules. A reduction of temperature to T=220K for TIP4P/2005 (below its melting point, that is, to within the supercooled regime) only produces qualitative changes, since the peak moves slightly to the left implying better structuring. This is notable both for bulk water and the hydrophilic systems and implies an improvement of the hydrogen bonds. Also, when we compare the results of the TIP4P/2005 model with that of TIP3P, both at T=300K, we find that the peaks for the former are located to the left as compared to that for the latter one. This is expected since, as already indicated, TIP3P has a much lower melting point and TIP4P/2005 works better in characterizing water structure.  

In turn, for the hydrophobic ($SAM-CH_3$) case, we find that a water molecule close to the surface is unable to retain full HB-coordination and, thus, is forced to resign one HB in the direction of the surface. At such tetrahedral site the $V_{4S}$ parameter shows that the water molecule feels a very low compensation from the surface, thus producing a sharp peak close to the mean thermal energy value ($k_B T$), a peak that has greatly decayed at a value of -6kJ/mol. This peak is very far in energy from the corresponding one for the water molecules at the hydrophilic $SAM-OH$ which occurs well below -12kJ/mol (a lower limit for hydrogen bond interactions\cite{v4,v4T2}). Additionally, when comparing the two water models and the two temperatures, the same trends are observed as in the case of the hydrophilic $SAM-OH$, but here only presenting small differences since the water-surface interactions are weak.

\begin{figure}[h!]
\resizebox{0.5\textwidth}{!}{%
\includegraphics{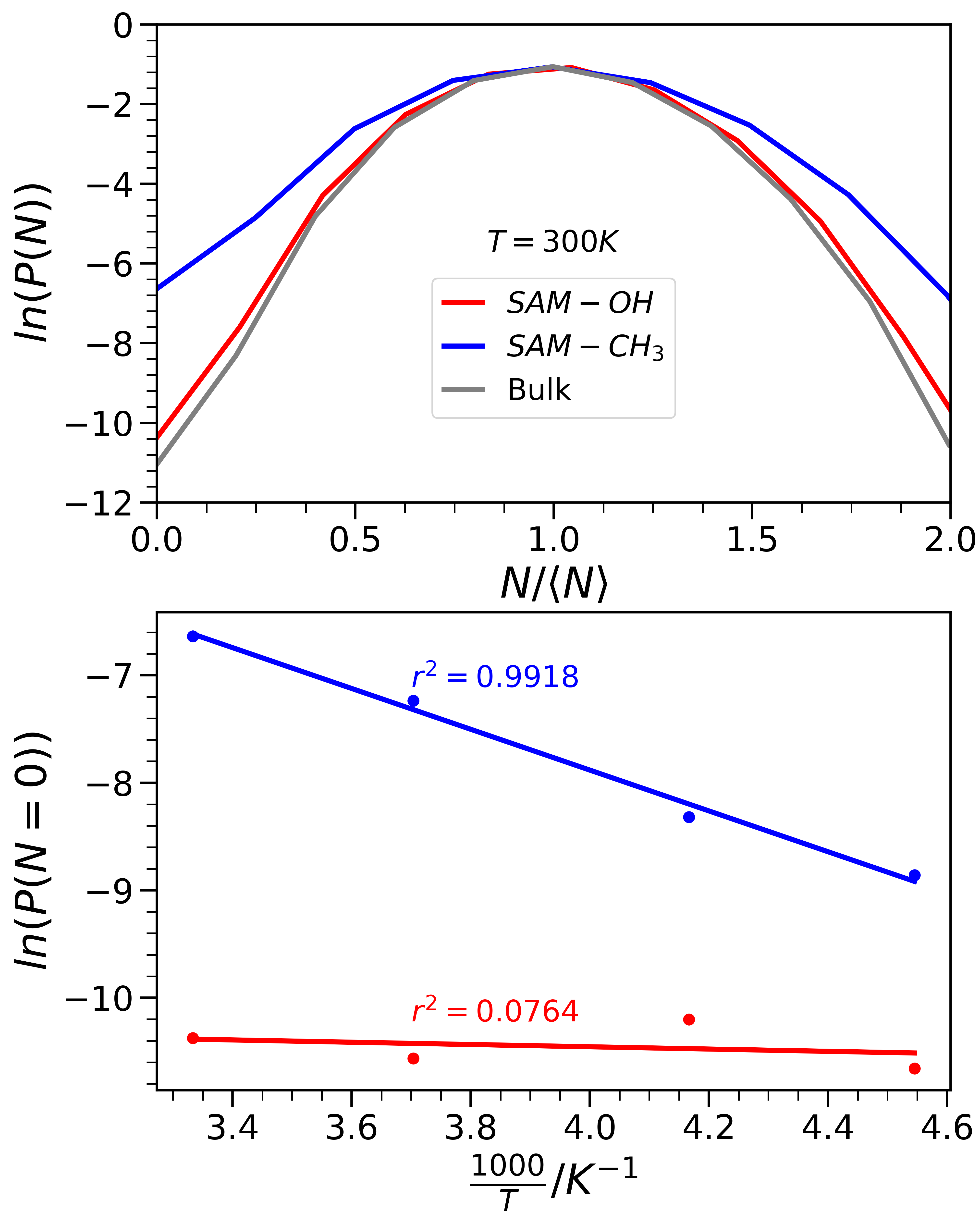}
}
\captionsetup{font=small}
\caption{Top: Probability of finding N water molecules within small spherical volumes tangent to the surface of both the hydrophobic and the hydrophilic SAMs (we also include the case for observation volumes within bulk water). P(N=0) gives the probability for cavity formation at such volume. A large value of P(N=0) for a surface system implies a hydrophobic behavior. Bottom: Behavior of the probability of cavity creation for the two SAMs at different temperatures. We note that $- k_B T \ln{P(N=0)}$ gives the work of cavity creation at the small observation volume.}
\label{fig4}
\end{figure}

The results exposed by Fig.~\ref{fig3} reveal the ability of the $V_{4S}$ index to accurately discriminate between hydrophobic and hydrophilic behaviors. For comparison, we note that an appropriate measure of hydrophobicity is provided by the extent of water density fluctuations and, thus, can be quantified by the probability of cavity creation in small volumes, P(N=0), where N is the number of water molecules inside the observation volume (we use spheres of radius r=3.3\AA, compatible with the size of a methane molecule, tangent to the surface)\cite{review_Garde}. In Fig.~\ref{fig4} we show the results for the SAMs under study. The $SAM-OH$ presents a low water vacating probability and, thus, typical hydrophilic behavior\cite{graphene-hydrophilic,review_Garde}. It is also evident that this value is very close to the water vacating probability in bulk water. In turn, the value for the $SAM-CH_3$ is much higher. This fact reveals a high dehydration propensity (or, equivalently, a low work for water removal or cavity creation) for the $SAM-CH_3$, thus implying hydrophobicity\cite{graphene-hydrophilic,review_Garde}. Thus, the results of the $V_{4S}$ index are nicely compatible with that of the water vacating probability. We also show a plot for $\ln P(N=0)$ as a function of inverse temperature for a range of temperatures studied for both SAMs. Consistently with the results of Fig.~\ref{fig3}, the hydropobic SAM shows a nice linear correlation, since the value of $V_{4S}$, the interaction that should be broken in order to desorb the water molecule, is roughly temperature-independent. In the case of the hydrophilic SAM this is not the case since the HB formed by the water molecule with the surface is improved as temperature decreases.

\begin{figure}[h!]
\resizebox{0.5\textwidth}{!}{%
\includegraphics{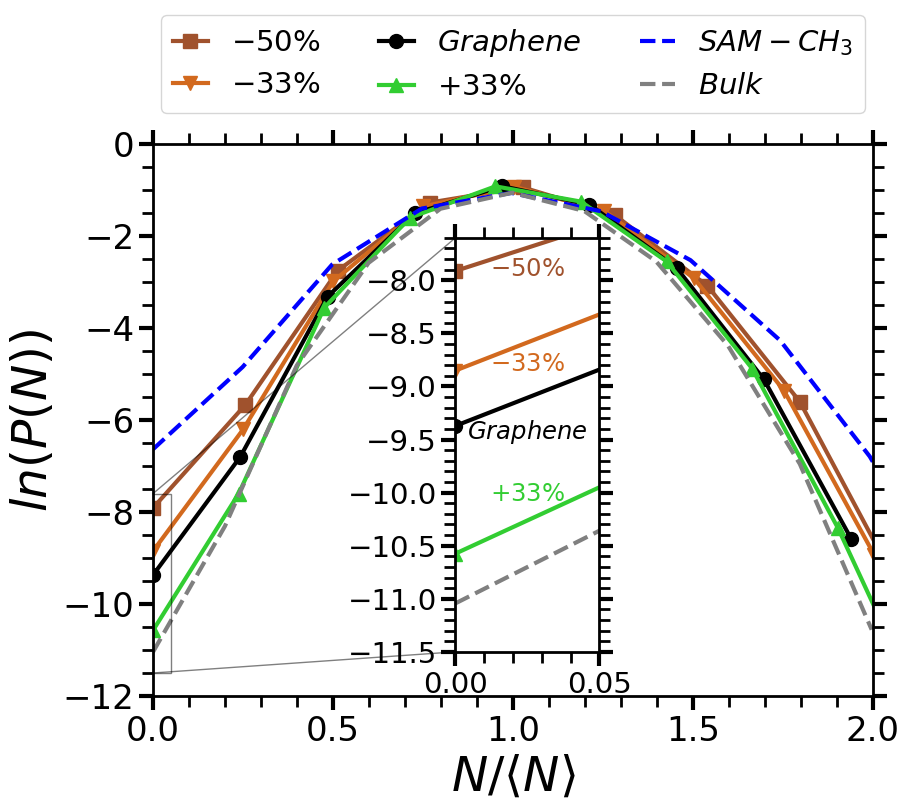}
}
\captionsetup{font=small}
\caption{Probability of finding N water molecules within small spherical volumes of radius r=3.3\AA\ tangent to the surface of a graphene surface and of graphene sheets with modified water-carbon interactions (33\% increase and 33\% and 50\% reduction). We also include the case for identical observation volumes within bulk water and tangent to the $SAM-CH_3$. The inset presents an expansion of a region of the graph to make the different probabilities of cavity creation, P(N=0), more clearly visible.}
\label{fig5}
\end{figure}

\begin{figure}[h!]
\resizebox{0.5\textwidth}{!}{%
\includegraphics{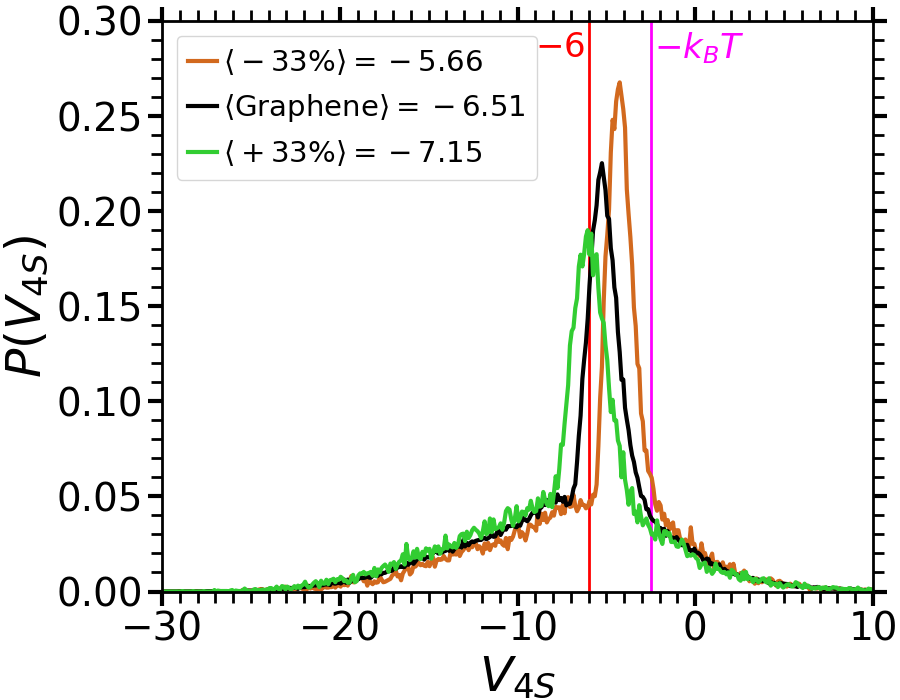}
}
\captionsetup{font=small}
\caption{Distributions of the $V_{4S}$ index for the water molecules close to a normal graphene sheet and to graphenes with (33\%) reduced and  (33\%) increased water-carbon interactions. We employ TIP4P/2005 water molecules and study cases at P=1 bar and T=300K. The mean values of the different distributions are indicated in the plot.}
\label{fig6}
\end{figure}

Another interesting system to study is graphene, which is expected to present a rather hydrophilic behavior\cite{graphene-hydrophilic,Fluid-Phase-Equilibria}. This system might also be amenable of producing either hydrophilic or hydrophobic behaviors by appropriately modifying the water-carbon interaction strength in the simulations and, thus, would be another convenient case where to test the ability of the $V_{4S}$ to act as a hydrophobicity measure. Indeed, the hydrophilic nature of graphene-based surfaces has been recognized both theoretically\cite{graphene-hydrophilic,Fluid-Phase-Equilibria} and by experimental water contact angle measurements over clean graphene-based surfaces (that showed that the rather hydrophobic-like values reported in previous studies had been in fact a result of aging and that arose from hydrocarbon adsorption from the air)\cite{graphene-exp}. Fig.~\ref{fig5} shows the water vacating probabilities within surface-tangent small radius spheres (P(N=0)) for a graphene sheet. We also provide results for modified graphene sheets were the water-carbon interactions (the $\epsilon_{C-O}$ value) have been modified by increasing them in 33\% with respect to the normal value and also by decreasing them in 33 and 50\%. An increase in the water-surface interactions is expected to make the surface more hydrophilic while a reduction in the interactions would decrease the hydrophilicity making the surface hydrophobic. The results for normal graphene exhibit a rather hydrophilic behavior since the P(N=0) value is much lower than that of the hydrophobic SAM, and relatively close to that of the hydrophilic SAM and bulk water. This fact is somehow intriguing, since it implies that the hydrophilic SAM and graphene present a similar dehydration propensity despite their different attraction intensities, suggesting that the work for water removal would change less quickly once overcoming certain hydrophilicity threshold. Indeed, graphene-like surfaces present a high carbon density and, thus, a neighboring water molecule would simultaneously interact with many carbon atoms, adding up several van der Waals interactions. This fact enables the establishment of a good hydrophilic wetting, making evident that nonpolar and hydrophobic are not synonyms. However, the total attraction that a water molecule experiences with a graphene-like surface is still far less intense as compared with a HB (roughly 35-40\% of the value that water attains at a $SAM-OH$ or in bulk conditions). Such a conundrum can be solved if we refer to the behavior of the $V_{4S}$ index shown in Fig.~\ref{fig6}, since the mean value of this indicator for the graphene surface is more negative than the mean value of $V_{4S}$ for the (high local density) D water molecules of the bulk water distribution ($V_{4S}$ around -6kJ/mol). This means that the water molecules in contact with a graphene-like surface receive at their HB-uncoordinated tetrahedral site a compensation that exceeds the value of the net interaction that a D molecule feels on average at its HB-lacking site in bulk conditions. Thus, to attain wettability and to reach hydrophilic-like behavior, it would not be necessary to recover full HB stabilization, as it happens at the $SAM-OH$ (LDL-like environments), but it would suffice to reach a threshold marked by the local energetic requirement of a typical D molecule within bulk HDL-like arrangements. In turn, if the interaction with any given surface implies a $V_{4S}$ value larger than $V_{4S}$=-6kJ/mol, the surface would exhibit a reduced wetting capability (reduced hydrophilicity), until attaining a neat hydrophobic behavior when the $V_{4S}$ value reaches that of the $SAM-CH_3$, close to $k_B T$ (and wetting becomes unfavorable as compared with a bulk D-state). Our results, thus, not only rationalize the hydrophobicity scale for these different surfaces, but also extends the validity of the two-liquids scenario to the hydration context by revealing the conditions for wettability.

When we increase the value of the water-carbon interactions in 33\%, the hydrophilicity of the system increases, as shown in Fig.~\ref{fig5}. A 33\% reduction of the interactions shows a small reduction in hydrophilicity while a more significant reduction (50\%) shifts the system to a behavior much more similar to that of the hydrophobic SAM. In Fig.~\ref{fig6} we also show the $V_{4S}$ distributions for the 33\% increase and reduction in the C-O interactions. While in the former case the mean value gets more negative, the latter case displays a mean value that crosses the threshold of the LDL-like environments (-6kJ/mol) which makes the system less favorable for water interaction and the behavior enters the hydrophobic regime.

\begin{figure}[h!]
\resizebox{0.5\textwidth}{!}{%
\includegraphics{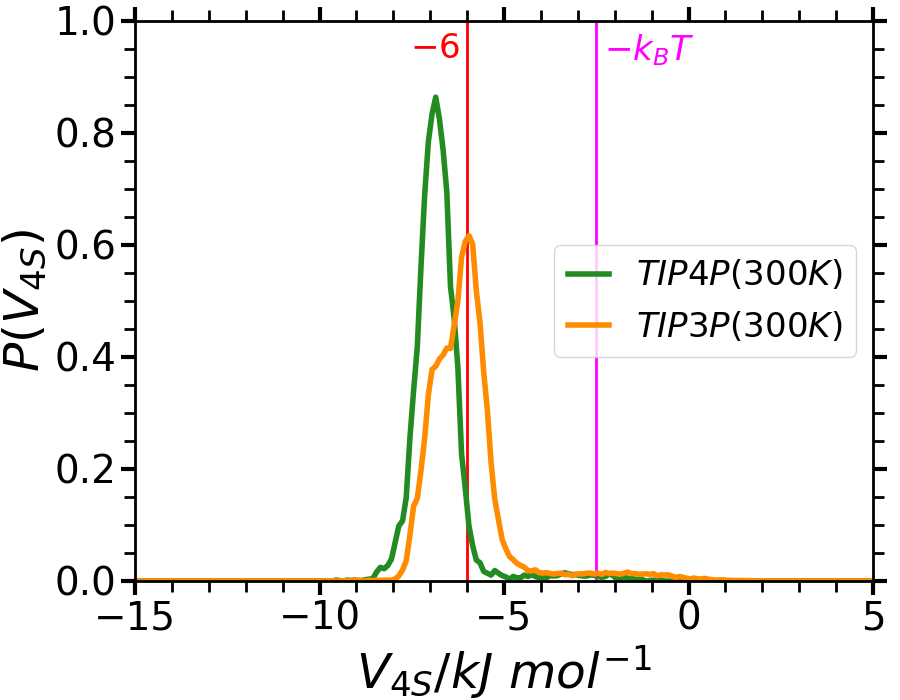}
}
\captionsetup{font=small}
\caption{Distributions of the $V_{4S}$ index for the water molecules inside a subnanometric section CNT. We show results for both TIP4P/2005 and TIP 3P water molecules, in both cases at T=300K.}
\label{fig7}
\end{figure}

Additionally, in Fig.~\ref{fig7} we display the $V_{4S}$ distribution for the water molecules in contact with the interior region of a narrow carbon nanotube of subnanometric section. The distribution of $V_{4S}$ for the CNT (as shown both for TIP3P and TIP4P/2005 at T=300K) displays a peak located within the hydrophilic range, at a value a bit more negative than the one corresponding to the normal graphene sheet. This is due to the fact that the CNT curvature enables the water molecules to establish a slightly better interaction with the dense carbon-network as compared to graphene. As expected, TIP4P/2005 yields a peak located a bit to the left as compared to TIP3P.

Finally, we note that a typical experimental measure of the hydrophobicity of a surface is given by the contact angle. The experimental contact angle for SAMs ended in OH is close to 10$^{\circ}$ (clearly hydrophilic) while for a SAM ended in methyls is larger than 100$^{\circ}$ which falls within the hydrophobic regime\cite{GardeSAMs}. For graphene, has also been shown to yield contact angles within the hydrophilic regime\cite{graphene-exp1,graphene-exp2}. It is experimentally difficult to measure contact angle of graphene on water since droplet deposition causes free-floating graphene to rupture. However, sophisticated experiments have determined that free-standing clean graphene is hydrophilic with a contact angle of 42$^{\circ}$ \cite{graphene-exp1,graphene-exp2}. Additionally, graphene on ice showed a contact angle of around 30$^{\circ}$ for a single graphene layer which slightly increased in 5$^{\circ}$ for graphene double layers on ice, while graphite yielded a hydrophilic value of around 60$^{\circ}$ \cite{graphene-exp1,graphene-exp2}. All these data are consistent with our results for the mean values of V4s of the corresponding systems where the -6kJ/mol would be taken as a hydrophilicity threshold.

\subsection {Nanoconfinement results: filling and drying carbon nanotubes}

\begin{figure*}[!]
\resizebox{0.9\textwidth}{!}{%
\includegraphics{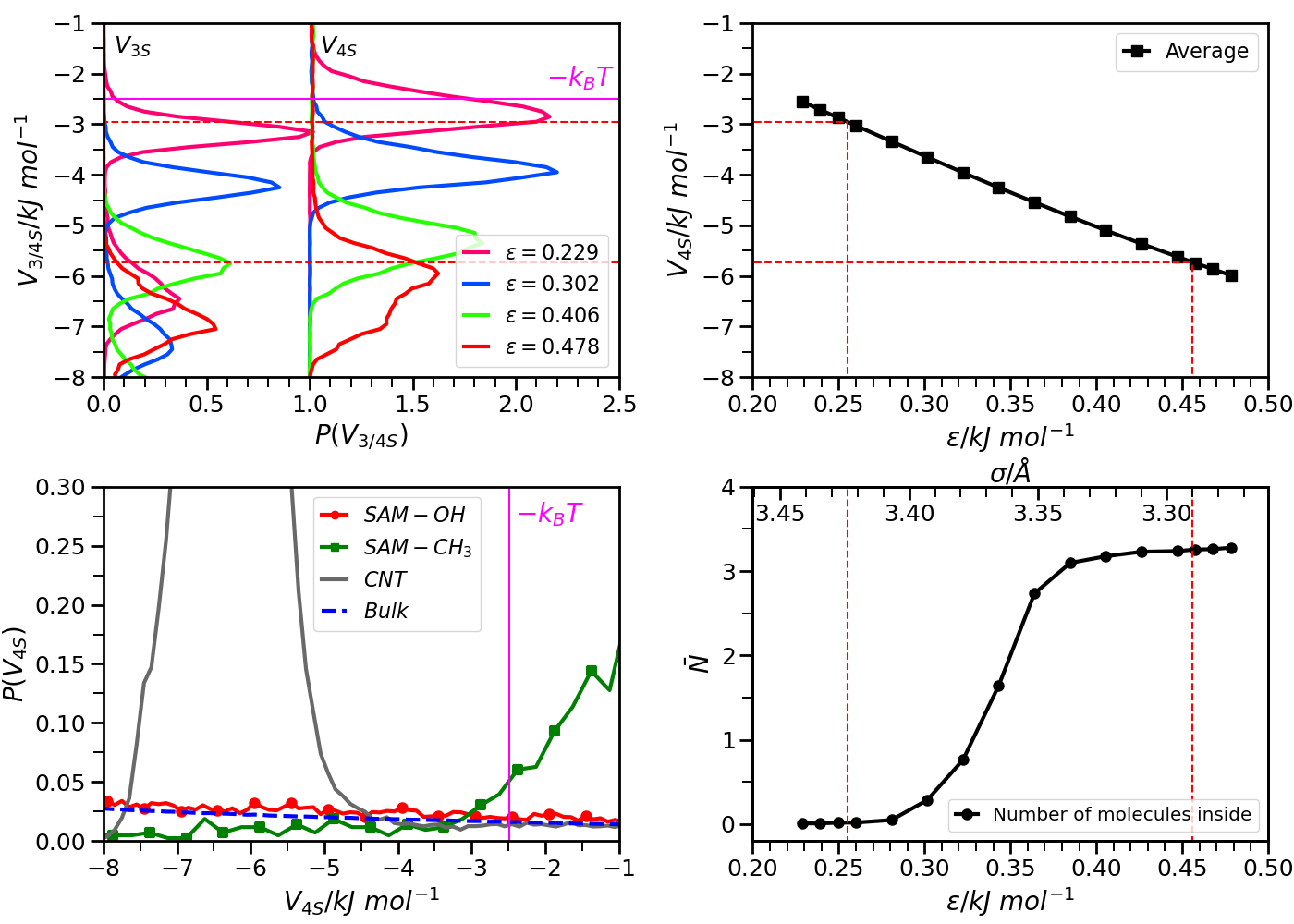}
}
\captionsetup{font=small}
\caption{Top left: Calculated distributions of $V_{4S}$ (solid lines) and $V_{3S}$ (dashed lines) for chosen values of  $\epsilon_{C-O}$; top right: we depict the corresponding average values of $V_{4S}$; bottom right: We show, for varying values of $\epsilon_{C-O}$, the mean number of molecules inside the CNT (we exclude from the calculation the regions at both CNT ends -2\AA\ length each- in order to avoid for possible border effects); bottom left: we include a detail of Fig.~\ref{fig3} for comparison. The red dashed lines indicate the estimated values for complete filling and complete drying.}
\label{fig8}
\end{figure*}

Concerning nanoconfinement, persistent penetration of small-diameter (subnanometric) carbon nanotubes by a one-dimensional arrangement of water molecules in simulations has long ago been deemed ``surprising''\cite{HummerCNT}, provided the fact that only a fraction of the energetic loss (resignation of HB coordination) could be recovered by water-wall attractions. Additionally, a partial reduction of such attractions promoted drying\cite{HummerCNT}. To demonstrate that such drying transitions can not only be explained by our approach but that also their range of occurrence can be quantitatively predicted, we studied a narrow CNT (see Methods), varying the Lennard-Jones parameters from $\epsilon_{C-O}=0.478369 \ kJ~{mol}^{-1}$ and $\sigma_{C-O}=3.27510$ \AA\ to $\epsilon_{C-O}=0.228720\ kJ~{mol}^{-1}$ and $\sigma_{C-O}=3.44154$ \AA, including the situations studied in\cite{HummerCNT}. Here we use the TIP3P water model precisely to compare with such work\cite{HummerCNT}. In Fig.~\ref{fig8} we provide the equilibrium average number of water molecules inside the CNT as we vary the Lennard-Jones parameters. Two evident regimes are detected close to each extreme $\epsilon$ values, as expected: completely filled for large $\epsilon$ and completely desorbed at low $\epsilon$. These two regimes are separated by a transition regime with a partial occupancy that results from an alternance of empty and (totally or partially) filled periods. In turn, starting from many CNT-filled configurations at $\epsilon_{C-O}=0.478369\ kJ~{mol}^{-1}$, we modified the Lennard-Jones parameters within the indicated linear regime, and we calculated the resulting $V_{4S}$ distributions. We also calculated the $V_3S$ distributions, since the water molecules inside the CNT resign two HBs (we note, however, that the values of $V_{4S}$ and $V_3S$ are  very similar, as expected). By so doing, we generated plots relating $\epsilon_{C-O}$ to mean $V_{4S}$ values, and we also included them in Fig.~\ref{fig8}. When we compare these plots, we find that starting from the highest $\epsilon_{C-O}$ value, the completely filled regime of the CNT ends at a value of  $\epsilon_{C-O}$ that produces a mean $V_{4S}$ value very close to -6kJ/mol, the one characteristic of the three-fold coordinated D water molecules in bulk conditions (cf. Fig.~\ref{fig2}). In turn, when we reach an $\epsilon_{C-O}$ value that yields a $V_{4S}$ close to -3kJ/mol (around the crossing between the $SAM-OH$ and $SAM-CH_3$, that is, the value which we have identified above as the beginning of the fully hydrophobic regime), the CNT becomes to be completely desorbed. Thus, it is evident that $V_{4S}$ is successful in quantitatively predicting filling/drying transitions in this nanoconfinement setting. Again, complete filling is compatible with a situation in which the CNT wall provides the water molecules with an environment equivalent to that of the high local density D molecules. It is not surprising, thus, that a full balance of the HB loss was not required for water in order to penetrate the CNT, but that the establishment of conditions compatible with the two-state scenario for water is demanded. Below such value of $\epsilon_{C-O}$, filling begins to be only partial, and the onset of complete drying occurs when the CNT walls are equivalent to the $SAM-CH_3$, as the value of $V_{4S}$ falls close to $k_B T$.

\section {Conclusions} We have shown that while liquid water presents a strong tendency to maximize the number of intermolecular hydrogen bonds (reducing energy by improving 4-fold coordination in the first-shell as the second neighbors are expanded), HB-coordination defects demand certain local compensation of the molecular charge at their HB-lacking position by second-shell contraction and orientation. This structural-energetic interplay gives rise to two preferred local molecular structures, as revealed by a novel structural metric, thus providing a detailed molecular rationalization of the two-liquids scenario. Moreover, we also show that the same molecular principle is operational at hydration and nanoconfinement conditions. In this sense, we explain both the unexpected high hydrophilicity of graphene-like systems (comparable to systems capable of stronger interactions with water and, thus, pointing to the possibility to define a hydrophilicity threshold) and the ``surprising'' penetration of water in small 
radius carbon nanotubes (where the water-wall interactions cannot fully recover the consequent loss in hydrogen bonding). In both cases, it suffices that the surface compensates the tetrahedral HB-lacking site at the level required by a high local density water molecule, thus revealing a main molecular rationale underlying hydration, wetability and nanoconfinement behavior for water and extending the validity of the two-liquids scenario to these realms. Furthermore, the structural metric we employ to test the molecular principle hereby revealed (which, at variance from the situation of previous structural indicators, is suitable for interfaces and nanoconfinement) enables us to quantitatively characterize hydrophobicity and to accurately predict water filling and drying transitions. Hence, it holds the promise to be of assistance in situations that involve different kinds of systems operating in aqueous environments within widespread contexts of biophysics and materials science.

{\bf Author contribution statement:} All authors contributed equally to this work.

{\bf Data Availability Statement:} Data sets generated during the current study are available from the corresponding author on reasonable request. Additionally, a programming code to implement $V_{4S}$ can be found in: https://github.com/nicolas-loubet/V4S

{\bf Acknowledments:} This work was possible thanks to public funding. The authors acknowledge support from CONICET and UNS.


\begin{thebibliography}{}
\bibitem{gallo_chemrev}
Gallo, P. et al. Water: A tale of two liquids. Chem. Rev. {\bf 116}, 7463-7500 (2016).
\bibitem{angell_02}
Angell, C. A. Liquid fragility and the glass transition in water and aqueous solutions. Chem. Rev. {\bf 102} 2627 (2002).
\bibitem{angell_04}
Angell, C. A. Amorphous water. Annu. Rev. Phys. Chem. {\bf 55} 559 (2004).
\bibitem{Debenedetti_book} P. G. Debenedetti, Metastable Liquids, Princeton University Press, Priceton, NJ,
(1996).
\bibitem{Tanaka00}
H. Tanaka, Simple physical model of liquid water. J. Chem. Phys. {\bf 112}, 799 (2000).
\bibitem{review_epje_1}
Gallo P. et al. Advances in the study of supercooled water Eur. Phys. J. E {\bf 44} 143 (2021). 
\bibitem{chaplin}
M. Chaplin, Do we underestimate the importance of water in cell biology? Nat. Rev. Mol. Cell Biol., {\bf 7} 861–866 (2006).
\bibitem{hassanali_chem_rev}
M-C. Bellissent-Funel, A. Hassanali, M. Havenith, R. Henchman, P. Pohl, F. Sterpone, D. van der Spoel, Y. Xu and A. E. Garcia. Water Determines the Structure and Dynamics of Proteins. Chem Rev {\bf 116}, 7673-97 (2016),  doi: 10.1021/acs.chemrev.5b00664. 
\bibitem{ball}
P. Ball. Water is an active matrix of life for cell and molecular biology. Proc. Natl. Acad. Sci. U.S.A. {\bf 114}, 13327–13335 (2017).
\bibitem{review_epje_2}
H. R. Corti et al. Structure and dynamics of nanoconfined water and aqueous solutions. Eur. Phys. J. E {\bf 44} 136 (2021).
\bibitem{debenedetti}
N. Giovambattista, P. J. Rossky, and P. G. Debenedetti. Computational Studies of Pressure, Temperature and Surface Effects on the Structure and Thermodynamics of Confined Water. Annu. Rev. Phys. Chem. {\bf 63}, 179-200 (2012).
\bibitem{berne}
C. Wang, B.J. Berne, R.A. Friesner. Ligand binding to protein-binding pockets with wet and dry regions. Proc. Natl. Acad. Sci. U.S.A. {\bf 108}, 1326-1330 (2011).
\bibitem{AFlibro}
A, Fernandez. Transformative Concepts for Drug Design: Target Wrapping. Springer Berlin, Heidelberg, ISBN: 978-3-642-11791-6 (2010); https://doi.org/10.1007/978-3-642-11792-3
\bibitem{water_m1}
E. Schulz, M. Frechero, G. A. Appignanesi and A. Fernández. {\it PLoS ONE} \textbf{5}, e12844 (2010). 
\bibitem{water_m4}
Accordino, S. R.; Malaspina, D. C.; Rodriguez Fris, J. A.; Alarc\'on, L. M.; Appignanesi, G. A. Temperature dependence of the structure of protein hydration water and the liquid-liquid transition. Phys. Rev. E {\bf 85}, 031503 (2012).
\bibitem{water_m8}
Accordino, S. R.; Malaspina, D. C.; Rodriguez Fris, J. A.; Appignanesi, G. A. Comment on ``Glass Transition in Biomolecules and the Liquid-Liquid Critical Point of Water". Phys. Rev. Lett. {\bf 106}, 029801 (2011).
\bibitem{marcia}
J. P. K. Abal and M. C. Barbosa. Water mobility in MoS2 nanopores: effects of the dipole–dipole interaction on the physics of fluid transport. Phys. Chem. Chem. Phys. {\bf 23}, 12075-12081 (2021).
\bibitem{bordin}
M. B. Le\~{a}o, J. R. Bordin, and C. F. de Matos. Specific surface area versus adsorptive capacity: an application view of 3D graphene-based materials for the removal of emerging water pollutants. Water Air Soil Pollut {\bf 234}, 136 (2023).
\bibitem{paolaV4}
G. Camisasca, L. Tenuzzo and P. Gallo.Protein hydration water: Focus on low density and high density local structures upon cooling. J. Mol. Liq. {\bf 370}, 120962 (2023).
\bibitem{graphene}
S. R. Accordino, J. M. Montes de Oca, J. A. Rodriguez Fris, G. A. Appignanesi. Hydrophilic behavior of graphene and graphene-based materials. J. Chem. Phys. {\bf 143}, 27 (2015)
\bibitem{cavities}
E. P. Schulz, L. M. Alarcón, G. A.  Appignanesi. Behavior of water in contact with model hydrophobic cavities and tunnels and carbon nanotubes. Eur. Phys. J. E {\bf 34} 114 (2011)
\bibitem{membrane}
L. M. Alarc\'on, M. de los Angeles Fr\'{i}as, M. A. Morini, M. B. Sierra, G. A. Appignanesi, E. A. Disalvo. Water Populations in Restricted Environments of Lipid Membrane Interphases. Eur. Phys. J. E {\bf 39}, 94 (2016).
\bibitem{martelli_franzese}
F. Martelli, J. Crain, and G. Franzese .Network Topology in Water Nanoconfined between Phospholipid Membranes. ACS Nano {\bf 14}, 8616-8623 (2020).
\bibitem{hydrophobicity1}
N. B. Rego and A. J. Patel. Understanding Hydrophobic Effects: Insights from Water Density Fluctuations. Annual Review of Condensed Matter Physics {\bf 13}, 303-324 (2022)
\bibitem{hydrophobicity2}
N. B. Rego, E. Xi and A. J. Patel. Identifying hydrophobic protein patches to inform protein interaction interfaces. Proc. Natl. Acad. Sci. USA {\bf 118} e2018234118 (2021)
\bibitem{hydrophobicity3}
I. Sinha, S. M. Cramer, H. S. Ashbaugh and S. Garde. Connecting Non-Gaussian Water Density Fluctuations to the Lengthscale Dependent Crossover in Hydrophobic Hydration. J. Phys. Chem. B {\bf 126}, 7604-7614 (2022).
\bibitem{bizzarri}
A. Bizzarri and S. Cannistraro. Molecular Dynamics of Water at the Protein-Solvent Interface. J. Phys. Chem. B {\bf 106}, 6617–6633 (2002).
\bibitem{short_time}
J. A. Rodriguez Fris, L.M. Alarc\'on and G. A. Appignanesi. Do short-time fluctuations predict the long-time dynamic heterogeneity in a supercooled liquid? Phys. Rev. E {\bf 76}, 011502 (2007).
\bibitem{identifying}
D. C. Malaspina, J. A. R. Fris, G. A. Appignanesi, F. Sciortino. Identifying a causal link between structure and dynamics in supercooled water. Europhys. Lett. {\bf 88}, 16003 (2009).
\bibitem{high-low}
J. M. Montes de Oca, et al. Structure and dynamics of high- and low-density water molecules in the liquid and supercooled regimes. Eur. Phys. J. E {\bf 39}, 124 (2016). https://doi.org/10.1140/epje/i2016-16124-4
\bibitem{proteins1} 
S. R. Accordino, J. A. Rodriguez Fris and G. A. Appignanesi. Wrapping effects within a proposed function-rescue strategy for the Y220C oncogenic mutation of protein p53. PLoS One {\bf 8}, e55123 (2013)
\bibitem{proteins2} 
S. R.  Accordino, et. al. Wrapping mimicking in drug‐like small molecules disruptive of protein–protein interfaces
Proteins: Structure, Function, and Bioinformatics {\bf 80}, 1755-1765 (2012); https://doi.org/10.1002/prot.24069
\bibitem{Poole92}
P. H. Poole, F. Sciortino, U. Essmann, and H. E. Stanley. Phase behaviour of metastable water,
Nature {\bf 360}, 324 (1992).
\bibitem{LLCP_T1}
F. Sciortino, I. Saika-Voivod, and P. H. Poole. Study of the ST2 model of water close to the liquid–liquid critical point.Phys. Chem. Chem. Phys. {\bf 13}, 19759–19764 (2011).
\bibitem{LLCP_T2}
J. C. Palmer, F. Martelli, Y. Liu, R. Car, A. Z. Panagiotopoulos, and P. G. Debenedetti. Metastable liquid–liquid transition in a molecular model of water. Nature {\bf 510}, 385 (2014).
\bibitem{Biddle}
J. W. Biddle, R. S. Singh, E. M. Sparano, F. Ricci, M. A. Gonzalez, C. Valeriani, J. L. F. Abascal, P. G. Debenedetti, M. A. Anisimov, and F. Caupin. Two-structure thermodynamics for the TIP4P/2005 model of water covering supercooled and deeply stretched regions. J. Chem. Phys. {\bf 146}, 034502 (2017).
\bibitem{LLCP_T3}
J. C. Palmer, P. H. Poole, F. Sciortino, and P. G. Debenedetti. Advances in computational studies of the liquid–liquid transition in water and water-like models. Chem. Rev. {\b f118}, 9129–9151 (2018).
\bibitem{Pablo-Francesco}
P. G. Debenedetti, F. Sciortino, and G. H. Zerze. Second critical point in two realistic models of water. Science {\bf 369}, 289 (2020)
\bibitem{LLCP_E1}
A. Nilsson and L. G. M. Pettersson, Perspective on the structure of liquid water, Chem. Phys. {\bf 389}, 1–34 (2011).
\bibitem{LLCP_E2}
K. H. Kim, A. Späh, H. Pathak, F. Perakis, D. Mariedahl, K. Amann-Winkel, J. A. Sellberg, J. H. Lee, S. Kim, J. Park et al., Maxima in the thermodynamic response and correlation functions of deeply supercooled water. Science {\bf 358}, 1589–1593 (2017).
\bibitem{LLCP_E3}
N. Stern, M. Seidl-Nigsch and T. Loerting. Evidence for high-density liquid water between 0.1 and 0.3 GPa near 150 K. Proc. Natl. Acad. Sci. U. S. A. {\bf 116}, 9191–9196 (2019).
\bibitem{Kim}
K. H. Kim, et al.. Experimental observation of the liquid-liquid transition in bulk supercooled water under pressure. Science {\bf 370}, 978 (2020).
\bibitem{epi}
A. Fernández. {\it Phys. Rev. Lett.} \textbf{108}, 188102 (2012).
\bibitem{debenedetti2}
N. Giovambattista, P. G. Debenedetti and P. J. Rossky. {\it Proc. Natl. Acad. Sci. U. S. A.} \textbf{106}, 15181 (2009).
\bibitem{review_Garde}
S. N. Jamadagni, R. Godawat and S. Garde. {\it Annu. Rev. Chem. Biomol. Eng.} \textbf{2}, 147 (2011).
\bibitem{graphene-exp}
Z. Li, et al. {\it Nature Materials} \textbf{12}, 925 (2013).
\bibitem{graphene-hydrophilic}
S. R. Accordino, J. M. Montes de Oca, J. A. Rodriguez Fris and G. A. Appignanesi. {\it J. Chem. Phys.} \textbf{143}, 154704 (2015).
\bibitem{Fluid-Phase-Equilibria}
L. M. Alarcón, J. M. Montes de Oca, S. R. Accordino, J. A. Rodriguez Fris and G. A. Appignanesi. {\it Fluid Phase Equilibria} \textbf{362}, 81 (2014).
\bibitem{HBPLOS}
C. A. Menéndez, S. R. Accordino, D. C. Gerbino and G. A. Appignanesi. {\it PLoS One} \textbf{11} (10), e0165767 (2016).
\bibitem{membranes}
L. M. Alarcón, et al. {\it Eur. Phys. J. E} \textbf{39}, 1-9 (2016).
\bibitem{HummerCNT}
G. Hummer, J. C. Rasaiah and J. P. Noworyta. {\it Nature} \textbf{414}, 188 (2001).
\bibitem{Giovambattista}
N. Giovambattista, P. G. Debenedetti, C. F. Lopez and P. J. Rossky. {\it Proc. Natl. Acad. Sci. U. S. A.} \textbf{105}, 2274 (2008).
\bibitem{v4s}
N. A. Loubet, A. R. Verde, G. A. Appignanesi. {\it J. Chem. Phys.} \textbf{160}, 144502 (2024).
\bibitem{v4}
J. M. Montes de Oca, F. Sciortino and G. A. Appignanesi. {\it J. Chem. Phys.} \textbf{152}, 244503 (2020).
\bibitem{v4T2}
N. A. Loubet, A. R. Verde, J. A. Lockhart, G. A. Appignanesi. {\it J. Chem. Phys.}  \textbf{159}, 064512 (2023); doi: 10.1063/5.0159060.
\bibitem{COMMENT-PRL}
S. R. Accordino, D. C. Malaspina, J. A. Rodríguez Fris and G. A. Appignanesi. {\it Phys. Rev. Lett.} \textbf{106}, 029801 (2011).
\bibitem{gromacs}
H. J. C. Berendsen, D. van der Spoel and R. van Drunen. {\it Comput. Phys. Commun.} \textbf{91}, 43 (1995).
\bibitem{berendsen}
H. J. C. Berendsen, J. P. M. Postma, W. F. van Gunsteren, A. DiNola and J. R. Haak. {\it J. Chem. Phys.} \textbf{81}, 3684–3690 (1984).
\bibitem{prbarostat}
M. Parrinello, A. Rahman. Crystal Structure and Pair Potentials: A Molecular-Dynamics Study. {\it Phys. Rev. Lett.} \textbf{45},  1196--1199 (1980). 
\bibitem{amber}
D. A. Case, et al., AMBER9, University of California, SanFrancisco, CA, 2006.
\bibitem{HDA-LDA} 
J. L. Finney, A. Hallbrucker, I. Kohl, A. K. Soper, and D. T. Bowron, Phys. Rev. Lett. \textbf{88}, 225503 (2002).
\bibitem{HDL}
J. M. Montes de Oca, S. R. Accordino, A. R. Verde, L. M. Alarcón and G. A. Appignanesi. {\it Phys. Rev. E} \textbf{99}, 062601 (2019).
\bibitem{GardeSAMs}
R. Godawat, S. N. Jamadagni and S. Garde. Proc. Natl. Acad. Sci USA (PNAS) \textbf{106} 15119–15124 (2009).
\bibitem{graphene-exp1}
L. A. Belyaeva, P. M. G. van Deursen, K. I. Barbetsea and G. F. Schneider. Adv Mater. \textbf{30} 1703274 (2018).
\bibitem{graphene-exp2}
A. V. Prydatko, L. A. Belyaeva, L. Jiang, L. et al. Nat. Commun. \textbf{9}, 4185 (2018).



\end{thebibliography}
\end{document}